\documentclass[aip,graphicx]{revtex4-2}
\usepackage{mathptmx}
\usepackage{graphicx}
\usepackage{float}

\begin{document}


\title{Evaluation of the AMOEBA force field for simulating metal halide perovskites in the solid state and in solution} 



\author{P.V.G.M. Rathnayake}
\affiliation{ARC Centre of Excellence in Exciton Science, School of Chemistry, University of Sydney, Sydney, New South Wales 2006, Australia}

\author{Stefano Bernardi}
\affiliation{ARC Centre of Excellence in Exciton Science, School of Chemistry, University of Sydney, Sydney, New South Wales 2006, Australia}

\author{Asaph Widmer-Cooper}
\affiliation{ARC Centre of Excellence in Exciton Science, School of Chemistry, University of Sydney, Sydney, New South Wales 2006, Australia}
\affiliation{The University of Sydney Nano Institute, University of Sydney, Sydney, New South Wales 2006, Australia}
\email[corresponding author: ]{asaph.widmer-cooper@sydney.edu.au}


\date{\today}

\begin{abstract}
In this work, we compare existing non-polarizable force fields developed to study the solid or solution phases of hybrid organic-inorganic halide perovskites with the AMOEBA polarizable force field. The aim is to test whether more computationally expensive polarizable force fields like AMOEBA offer better transferability between solution and solid phases, with the ultimate goal being the study of crystal nucleation, growth and other interfacial phenomena involving these ionic compounds. In the context of hybrid perovskites, AMOEBA force field parameters already exist for several elements in solution and we decided to leave them unchanged and to only parameterize the missing ones (Pb\textsuperscript{2+} and CH\textsubscript{3}NH\textsubscript{3}\textsuperscript{+} ions) in order to maximise transferability and avoid over-fitting to the specific examples studied here. Overall, we find that AMOEBA yields accurate hydration free energies (within 5\%) for typical ionic species while showing the correct ordering of stability for the different crystal polymorphs of CsPbI\textsubscript{3} and CH\textsubscript{3}NH\textsubscript{3}PbI\textsubscript{3}. While the existing parameters do not  accurately reproduce all transition temperatures and lattice parameters, AMOEBA offers better transferability between solution and solid states than existing non-polarizable force fields. 
\end{abstract}


\maketitle 

\section{Introduction}

Metal halide perovskites show promising potential as an efficient and cheap alternative for conventional solar cell materials. Since their first emergence in 2009, the power conversion efficiency has reached above 20\% within a decade\cite{kojima2009organometal,solarcellefficiencytables53}. A typical metal halide perovskite has an ABX\textsubscript{3} composition where A is an inorganic cation (Cs\textsuperscript{+}) or an organic cation (methylammonium ion, ethylammonium ion, formamidinium ion, etc...), B is an inorganic cation (Pb\textsuperscript{2+}, Sn\textsuperscript{2+}) and X is an anion (Cl\textsuperscript{-}, Br\textsuperscript{-}, I\textsuperscript{-}). Their low production cost, earth-abundant precursors, tunable bandgaps, and high power conversion efficiency, make these materials attractive for a range of applications \cite{solar-cell-efficiency-tables54, castelli2014bandgap, zhao2018thermodynamically, wang2016tunable}. However, further development is necessary before perovskite solar cells (PSCs) can be widely adopted by industry. For instance, PSCs are not as stable in humid environmental conditions as conventional “silicon-based” solar cells and it is not currently possible to print efficient perovskite solar cells at scale\cite{han2015degradation, shirayama2016degradation}. 
Even though a wide variety of specific fabrication methods exist to manufacture metal halide perovskites with desired properties and composition,  \cite{tidhar2014crystallization, saidaminov2015high, stranks2015formation, foley2017controlling, costa2017deposition, zhang2017centimeter} we still lack a deep understanding of how perovskite crystals form at the molecular level and of how to influence this process. 

Computational modelling can be a powerful tool to study the behaviour of materials at the molecular level. Numerous studies have investigated the structure and dynamics of metal halide perovskites in the solid state using ab-initio techniques \cite{carignano2015thermal, huang2018intrinsic}. However, due to the high computational cost of these methods, the accessible time is limited to 10s of picoseconds. 
This time is inadequate to observe activated processes like nucleation and growth, which typically require timescales that are many orders of magnitude longer \cite{hu2017situ,zhou2015microstructures}. Classical force fields, coupled with advanced sampling methods designed to drive the system toward specific physical states, can help bridge this gap. To date, several classical force fields have been specifically developed for hybrid perovskites\cite{handley2017new, hata2017development, gutierrez2015molecular, caddeo2017collective, bischak2019liquid}. Almost all classical force fields however, use point charge electrostatic interactions, which limits their transferability between different perovskites and especially between the solid state and solution. Polarizable force fields partly address this issue by incorporating many-body effects arising from electronic polarization. 

In this work, we used the AMOEBA polarizable force field to simulate CsPbI\textsubscript{3} and CH\textsubscript{3}NH\textsubscript{3}PbI\textsubscript{3} perovskites and their precursors in solution. Existing parameters were kept untouched, parameterizing only those elements missing from the available force field library. This was done in order to avoid overfitting and to preserve transferability. 
Solid and solution state properties of CH\textsubscript{3}NH\textsubscript{3}PbI\textsubscript{3} perovskites were also examined using two existing non-polarizable force fields: MYP1\cite{caddeo2017collective}, and the force field developed by Gutierrez-Sevillano \textit{et al}\cite{gutierrez2015molecular}. These force fields were parameterized to describe both solid and solution state properties of CH\textsubscript{3}NH\textsubscript{3}PbI\textsubscript{3}, whereas the other existing force fields were developed only with the solid state in mind\cite{handley2017new, hata2017development, caddeo2017collective, bischak2019liquid}.

\section{Methods}
\subsection{AMOEBA force field Parameterization}
The AMOEBA force field includes bond stretching ($U\textsubscript{bond}$), angle bending ($U\textsubscript{angle}$), bond-angle cross term ($U\textsubscript{bond-angle}$), out of plane bending ($U\textsubscript{oop}$), and torsional rotation (U\textsubscript{torsion}) terms. Non-bonding terms include van der Waals ($U\textsubscript{vdW}$) and electrostatic contributions. Electrostatic interactions include contributions from both permanent ($U\textsubscript{ele-permanent}$) and induced multipoles ($U\textsubscript{ele-induced}$). The total potential, U\textsubscript{AMOEBA}, is given by equation \ref{eqn:amoeba}.
\begin{equation} \label{eqn:amoeba}
    U\textsubscript{AMOEBA} = U\textsubscript{bond} + U\textsubscript{angle} + U\textsubscript{bond-angle} + U\textsubscript{oop} + U\textsubscript{torsion}  + U\textsubscript{vdW} + U\textsubscript{ele-permanent} + U\textsubscript{ele-induced}
\end{equation}

AMOEBA comes with a library of parameters for small organic molecules\cite{ren2011polarizable, shi2018dft}, ions\cite{wu2010polarizable, xia2017polarizable, piquemal2006towards}, amino acids\cite{shi2013polarizable}, lipids\cite{chu2018polarizable}, nucleic acids\cite{zhang2018amoeba} and water\cite{ren2003polarizable, wang2013systematic, laury2015revised}. Parameters for Pb\textsuperscript{2+} and methylammonium ions needed for hybrid perovskites are missing. However, a clear guideline on how to parameterize new elements/compounds is provided by the authors\cite{ren2011polarizable}. 

Since atomic ions like Pb\textsuperscript{2+} are spherically symmetric, no higher order multipole components are needed. We set the Pb\textsuperscript{2+} monopole to +2 to preserve its full ionic charge. The polarizability was obtained from ab-initio calculations at the MP2/cc-pVQZ-PP level using the Gaussian16 software package\cite{g16}. 
A frozen effective core potential\cite{peterson2003systematically} was used for these calculations. The external basis set was obtained from the EMSL Basis Set Exchange database\cite{schuchardt2007basis}. The van der Waals (vdW) parameters $R_{min}$ (minimum energy distance ) and $\epsilon$ (potential well depth) cannot be directly extracted from ab-initio calculations. Instead, the AMOEBA vdW parameters for Pb\textsuperscript{2+} were obtained by fitting to the ab-initio binding potential between Pb\textsuperscript{2+} and H\textsubscript{2}O after correcting for basis set superposition error. Mixed basis sets of cc-pVQZ-PP and aug-cc-pVQZ were used for Pb\textsuperscript{2+} and water, respectively. 

AMOEBA uses the Thole screening function\cite{van1998molecular, thole1981molecular} $\rho$ to avoid the polarization catastrophe by smearing the atomic multipole moment at short interatomic distances. This is defined in equation \ref{eqn:thole}, where $u=r_{ij}/(\alpha_i\alpha_j)^{1/6}$, $r_{ij}$ is the interatomic distance, $\alpha_i$ is the polarizability of atom $i$, and $a$ is the dimensionless Thole damping factor (which controls the damping strength). AMOEBA uses the universal damping factor 0.39. However, multivalent ions typically require a broader charge distribution and thus a smaller $a$. For example, such modifications had been reported for Ca\textsuperscript{2+}, Mg\textsuperscript{2+} and Zn\textsuperscript{2+} ions\cite{piquemal2006towards, wu2010polarizable}. In our fitting procedure, the Thole parameter for Pb\textsuperscript{2+} was initially kept at its default value of 0.39. Once good vdW parameters were found, the solid and liquid phase properties were checked. As the default Thole parameter resulted in an unstable crystal it was reduced to produce a stable solid and decrease the 3D-orthorhombic to cubic phase transition temperature towards the experimental value. The vdW parameters were subsequently refitted with the new Thole damping factor. 

\begin{equation}\label{eqn:thole}
    \rho=\frac{3a}{4\pi}exp(-au^3)
\end{equation}

The methylammonium ion was manually parameterized following standard methods\cite{ren2011polarizable}. All new AMOEBA parameters can be found in the supplementary data. Unaltered AMOEBA09 parameters were used for I\textsuperscript{-}, Br\textsuperscript{-}, Cl\textsuperscript{-}, Cs\textsuperscript{+}, water and dimethylformamide (DMF).

\subsection{Non-polarizable force fields}
The MYP family of non-polarizable force fields has shown good agreement with a range of experimental properties of CH\textsubscript{3}NH\textsubscript{3}PbI\textsubscript{3} perovskites. MYP0\cite{mattoni2015methylammonium} was specifically developed to describe solid state properties, while its successor, MYP1\cite{caddeo2017collective}, included water interactions to study layer-wise degradation while leaving solid state properties unaltered\cite{caddeo2017collective}. As we are interested in force fields that can describe both solution and solid-state properties, we focus on the MYP1 variant here. MYP1 describes the hybrid perovskite as the sum of organic-organic (U\textsubscript{OO}), inorganic-inorganic (U\textsubscript{II}) and organic-inorganic (U\textsubscript{OI}) interactions. The organic-organic interactions follow the generalised AMBER force field (GAFF) structure\cite{wang2004gaff} (equation \ref{eqn:uoo1}-\ref{eqn:uoo3}), the inorganic-inorganic interactions are described by a Buckingham-Coulomb potential (equation \ref{eqn:uii}), while the hybrid organic-inorganic interactions are described as the sum of Buckingham, Coulombic and Lennard-Jones terms (equation \ref{eqn:io}).
\begin{eqnarray} \label{eqn:uoo}
    U\textsubscript{OO} &=& U\textsubscript{bonded} + U\textsubscript{non-bonded}  \label{eqn:uoo1}\\
    U\textsubscript{bonded} &=& U\textsubscript{bonds} + U\textsubscript{angles} + U\textsubscript{dihedrals} \label{eqn:uoo2} \\
    U\textsubscript{non-bonded} &=&  U\textsubscript{LJ} + U\textsubscript{Coul} \label{eqn:uoo3}
\end{eqnarray}
\begin{eqnarray} 
    U\textsubscript{II} &=& U\textsubscript{Buck} + U\textsubscript{Coul} \label{eqn:uii}
\end{eqnarray}
\begin{eqnarray} 
        U\textsubscript{IO} &=& U\textsubscript{Buck} + U\textsubscript{LJ} + U\textsubscript{Coul} \label{eqn:io}
\end{eqnarray}

Gutierrez-Sevillano and co-workers\cite{gutierrez2015molecular} used a different force field to study the formation of pre-critical CH\textsubscript{3}NH\textsubscript{3}PbI\textsubscript{3} clusters from precursor ions, Pb\textsuperscript{2+}, I\textsuperscript{-} and CH\textsubscript{3}NH\textsubscript{3}\textsuperscript{+} in water, pentane and water-pentane binary mixtures. Specifically, they used a rigid all-atom methylammonium ion model with atomic positions taken from the open chemistry database and non-bonding parameters taken from the AMBER force field\cite{wang2004development}. For water, they used the all-atom SPC/E model, and for the organic solvents parameters were taken from the TraPPE united-atom force field. Pb\textsuperscript{2+} and I\textsuperscript{-} parameters were selected from other studies based on their ability to yield good hydration free energies. The basic potential form is similar to a simplified version of the AMBER force field (equations \ref{eqn:gaff1} and \ref{eqn:gaff2}). 
\begin{eqnarray}
    U\textsubscript{bonded} &=& U\textsubscript{stretch} + U\textsubscript{bend} + U\textsubscript{torsion}  \label{eqn:gaff1} \\
    U\textsubscript{non-bonded} &=&  U\textsubscript{LJ} + U\textsubscript{Coul}  \label{eqn:gaff2}
\end{eqnarray}

All non-polarizable force field calculations were performed using LAMMPS\cite{lammps, plimpton1995fast}.

\subsection{Solution state simulations}
Hydration Free Energies (HFEs) were calculated for Pb\textsuperscript{2+}, CH\textsubscript{3}NH\textsubscript{3}\textsuperscript{+} and halide ions (Cl\textsuperscript{-}, Br\textsuperscript{-}, I\textsuperscript{-}) solvated in a cubic box containing 512 water molecules. 

\subsubsection{Non-polarizable models}
Prior to the HFE calculation, the ion-water box was equilibrated for 1 ns at constant temperature (300 K) and pressure (1 atm), using a Nose-Hoover thermostat and barostat. The HFEs were determined using the finite-difference thermodynamic integration (FDTI) method\cite{mezei1987finite} in user package FEP in LAMMPS. Soft-core potentials were applied to ion-water pairs to avoid singularities when the coupling parameter $\lambda$ approaches zero\cite{beutler1994avoiding}. The same water models as in the original force fields, i.e. TIP3P for MYP1 and SPC/E for Gutierrez-Sevillano's potential, were used. A one step transformation (ion discharging and decoupling) was performed during a 10 ns simulation using a 1 fs timestep and a $\lambda$ increment of 10\textsuperscript{-4}. Volume changes were not considered in the free energy calculation and statistics were collected every 1 ps (yielding a total of 10\textsuperscript{4} data points). Both ion growth and disappearance were studied.

\subsubsection{AMOEBA polarizable model}
The HFEs for the AMOEBA force field were calculated in Tinker-OpenMM\cite{harger2017tinker} using the Bennett Acceptance Ratio method (BAR) and the conventional thermodynamic cycle: ion discharge in solution, decouple in solution, and ion discharge in vacuum. The hydration free energy can be calculated via equation \ref{eqn3}, where $\Delta G_{discharge(vac)}$ is the free energy change with the gradual charge removal of the ion in  vacuum,
$\Delta G_{discharge(water)}$ is the free energy change with gradual charge removal of the ion in water, and 
$\Delta G_{decouple(water)}$ is the free energy change to turn off the vdW interaction between the ion and water molecules. Note that it is important to perform the discharging step first to avoid singularities between electrostatic charges due to the absence of vdW interactions.  
\begin{equation} \label{eqn3}
    \Delta G_{hydration} = \Delta G_{discharge (vac)} - \Delta G _{discharge (water)} - \Delta G _{decouple (water)}
\end{equation}

The ion-water system was minimized using the steepest-descent algorithm. The system was then equilibrated at 300 K at constant volume for 100 ps and afterwards at a constant pressure of 1 atm (same temperature) for 2 ns to determine the average volume. This equilibrated structure was used for all subsequent $\lambda$ simulations. Each $\lambda$ window was run for 1 ns at 300 K using the average volume determined earlier. The first 200 ps of each window was discarded and only the last 800 ps used for the BAR analysis. Fifteen $\lambda$ windows were found to be adequate to converge the calculations ($\lambda$ = 0.0, 0.1, 0.2, 0.3, 0.4, 0.45, 0.5, 0.55, 0.6, 0.65, 0.7, 0.75, 0.8, 0.85, 0.9, 1.0).

\subsubsection{Potential of Mean Force Calculations}
Constrained molecular dynamics simulations were used to calculate the potential of mean force (PMF) as a function of the lead-halide ion separation. Calculations were performed for Pb-X (X=Cl\textsuperscript{-}, I\textsuperscript{-}) ion pairs in DMF. A simulation box filled with 500 DMF molecules was equilibrated at 300 K and 1 atm using an Andersen thermostat and a Monte Carlo isotropic barostat in OpenMM. The short range cutoff was set to 1.2 nm, mutual polarisation was imposed with a 10\textsuperscript{-5} tolerance and the Particle Mesh Ewald (PME) method was used to compute electrostatic interactions with periodic boundary conditions. Simulations were performed at Pb-X separations (in Angstroms) of 2.0, 2.6, 2.8-4.6 (in increments of 0.1), 4.8-6.8 (in increments of 0.2), 7.0-12.5 (in increments of 0.5) and 13-16 (in increments of 1). When adding the second halide ion, a loose harmonic spring (K=1.2 kcal/mol/{\AA}\textsuperscript{2}) was used to keep the first ion bonded to lead, with the equilibrium bond lengths taken from crystallographic data; 3.12 {\AA} and 2.84 {\AA} for Pb-I and Pb-Cl, respectively. A simulation of 1 ns was performed for each value of the reaction coordinate, with the last 600 ps used to calculate the PMF.

 The free energy difference, $\Delta F(z_s)$, between a state where the halide ion is positioned at $z_s$ and a reference state where the halide ion is at $z_0$ is given by equation \ref{eq:pmf}, where $f_z(z_s^{'})$ is the $z$ component of the force exerted by the solvent and lead ion on the halide ion at the position $z_s^{'}$. The average forces at each step were then interpolated and integrated to obtain the PMF\cite{dang1999computer}. 
\begin{equation} \label{eq:pmf}
    \Delta F(z_s) = F(z_s)-F_0 = - \int_{z_0}^{z_s}\left \langle f_z (z_s^{'} ) \right \rangle dz_s^{'}
\end{equation}

\subsection{Solid State Simulations}
Both CsPbI\textsubscript{3} and CH\textsubscript{3}NH\textsubscript{3}PbI\textsubscript{3} solid phases were examined with AMOEBA using OpenMM with GPU acceleration. All OpenMM simulations were carried out with a Verlet Integrator, Andersen thermostat and a Monte Carlo anisotropic barostat. 1 atm pressure was maintained for all simulations. Non-bonding cutoffs were set to 1.2 nm. Periodic boundary conditions were used with the Particle Mesh Ewald (PME) method. Mutual polarisation was imposed with 10\textsuperscript{-5} tolerance. 

For CsPbI\textsubscript{3}, 1D and 3D orthorhombic phases were heated from 200-900 K in increments of 50 K. Extra steps were added between 500-700 K with increments of 10 K to increase the resolution in the phase transition regions. All steps were equilibrated for 2 ns, with subsequent steps continued from the previous step. The same procedure was used for CH\textsubscript{3}NH\textsubscript{3}PbI\textsubscript{3}, with the temperature range set to 100-450 K and 10 K increments used from 150-360 K. 

A similar procedure was used to study the solid phases of CH\textsubscript{3}NH\textsubscript{3}PbI\textsubscript{3} with the MYP1 point charge model using LAMMPS. These simulations were run at 1 atm pressure using a 0.5 fs timestep.

\section{Results and Discussion}
\subsection{Solution Properties}
\subsubsection{The Hydration Free Energy (HFE)}
The HFE is a key property to validate a force field in solution. We calculated the HFEs of the precursor ions with both polarizable and  non-polarizable force fields (table \ref{tab:table1}). Standard state corrections were applied to the calculated HFEs \cite{grossfield2003ion} (and also to the experimental results of Schmid \textit{et al.}, which were not determined under standard state conditions). The simulation box size was chosen to minimize finite size effects and ensure good convergence to bulk values \cite{grossfield2003ion}. HFEs calculated with the polarizable AMOEBA force field show very good agreement with experimental data (within $5\%$ except for methylammonium). In contrast, the MYP1 force field consistently underestimates the hydration free energies by 50\% or more. The discharging process in water is highly sensitive to the ionic charge and $\Delta G_{discharge(water)}$ dominates the total HFE. Due to this, force fields with reduced ionic charges, such as MYP1 generally exhibit HFEs that are too low. The force field of Gutierrez-Sevillano and co-workers\cite{gutierrez2015molecular} uses full charges and yields more realistic HFEs, but the HFE is still ~10\% too low for Pb\textsuperscript{2+}. Solvation free energies in organic solvents (e.g. DMF and DMSO) were not calculated as, to our knowledge, no reliable experimental data exists to compare such values with.

\begin{table}
\centering
\begin{tabular}{ p{1cm} p{2cm} p{2cm} p{2cm} p{2cm} p{2cm} }
 \hline
 \multicolumn{6}{c}{Hydration Free Energy (kcal/mol)} \\
 \hline
 & \multicolumn{2}{c}{Experimental data} & \multicolumn{3}{c}{Force Field data} \\
 \hline
 & Schmid \textit{et al.} \cite{schmid2000new} & Marcus \textit{et al.} \cite{marcus1987thermodynamics, marcus1994simple} & AMOEBA & MYP1 & Gutierrez-Sevillano \textit{et al} \\
 \hline
 Cl\textsuperscript{-}  & -87.3  & -81.3   & -83.8   &          &          \\
 Br\textsuperscript{-}  & -80.7  & -75.3   & -75.8   &          &          \\
 I\textsuperscript{-}   & -72.4  & -65.7   & -66.9   & -32.8   & -64.7   \\
 Cs\textsuperscript{+}  & -58.6  & -59.8   & -56.6   &          &          \\
 Pb\textsuperscript{2+} &        & -340.6  & -338.7  & -134.5  & -316.8  \\
 CH\textsubscript{3}NH\textsubscript{3}\textsuperscript{+}    &         & -70.5   & -60.7   & -30.4   &          \\
 \hline
\end{tabular}
\caption{\label{tab:table1}Experimental and Calculated Hydration Free Energies from different sources and force fields.}
\end{table}

\subsubsection{Halide coordination to Pb\textsuperscript{2+} in solution}
\begin{figure}
    \includegraphics[scale=0.6]{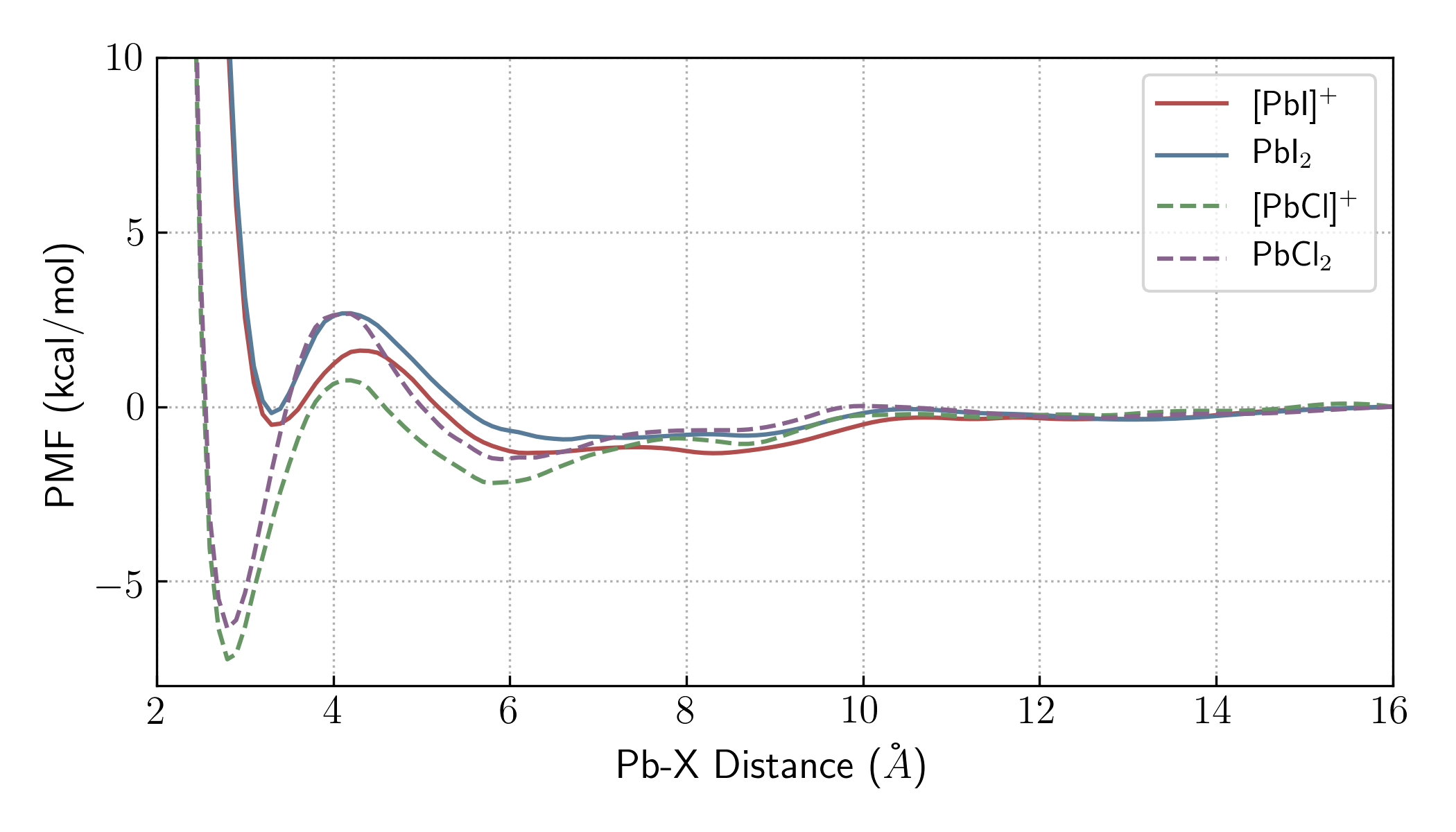}
    \caption{Potentials of mean force (PMF) for the formation of [PbX]\textsuperscript{+} and PbX\textsubscript{2} species in DMF at 300 K upon successive halide addition.}
    \label{fig:pmf}
\end{figure}

The choice of solvent is crucial in controlling the morphology of hybrid perovskites\cite{zhao2014solution}, and DMF is a widely used solvent in perovskite preparation. In the absence of reliable solvation free energies in DMF, we have characterized the initial steps in the formation of pre-nucleation clusters in this solvent by calculating the change in free energy as a function of the lead-halide separation for several complexes at 300 K. The resulting PMFs are shown in figure \ref{fig:pmf} for both I\textsuperscript{-} and Cl\textsuperscript{-} additions. All of the potentials exhibit a broad outer minimum, a barrier, and a narrow inner minimum where the halide is directly coordinated to Pb\textsuperscript{2+}. The outer minima correspond to solvent separated states, while the barriers near 4 {\AA} are due to displacement of DMF molecules from the inner coordination shell as the halides approach. This behavior is similar to the ion pairing free energy in water for Ca-SO\textsubscript{4} (also calculated using AMOEBA).\cite{byrne2017computational} In our case, the PMFs indicate that chloride prefers to form contact ion pairs with Pb\textsuperscript{2+} while iodide prefers to form solvent-separated states, which is consistent with the experimental observation that PbI\textsubscript{2} has greater solubility than PbCl\textsubscript{2} in DMF at room temperature\cite{synnott1969chloride, gaizer1967reaction}. Note that this is opposite to the solubility order of these salts in water, which indicates that our extension to the AMOEBA force field may be sufficiently accurate to study the formation of metal halide perovskites in commonly used solvents such as DMF. 

\subsection{Structural Properties}
\subsubsection{CsPbI\textsubscript{3}}
Experimentally, upon heating, CsPbI\textsubscript{3} transforms from a 1D-orthorhombic polymorph into a photo-active cubic structure, (a\textsuperscript{0}a\textsuperscript{0}a\textsuperscript{0}) tilting pattern in Glazer notation\cite{glazer1972classification}, near 600 K\cite{marronnier2018anharmonicity, trots2008high}. In-situ powder diffraction indicates that this transition is first order\cite{trots2008high}. Upon cooling from the cubic phase, metastable tetragonal (a\textsuperscript{0}a\textsuperscript{0}c\textsuperscript{+}) and 3D-orthorhombic (a\textsuperscript{-}a\textsuperscript{-}a\textsuperscript{+}) polymorphs can also form\cite{marronnier2018anharmonicity,sutton2018cubic}.

\begin{figure}[H]
    \centering
    \includegraphics[scale=0.5]{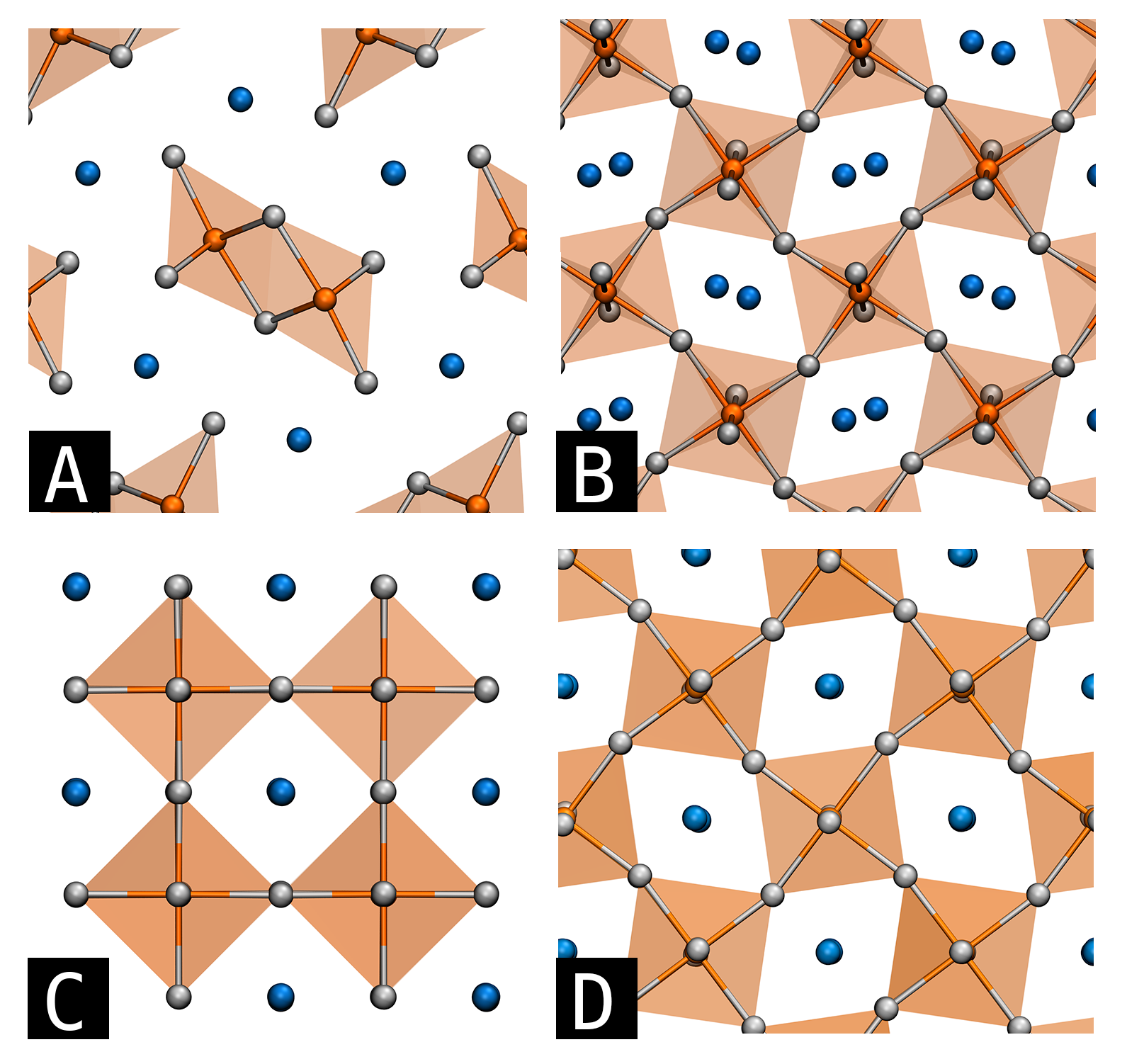}
    \caption{Averaged atomic positions of CsPbI\textsubscript{3}  polymorphs from simulations with the AMOEBA force field. Pb, I and Cs atoms are colored in orange, grey and blue, respectively. (A) 1D-orthorhombic phase at 100 K, (B) 3D-orthorhombic phase at  450 K (a\textsuperscript{-}a\textsuperscript{-}c\textsuperscript{+}), (C) cubic phase at 700 K (a\textsuperscript{0}a\textsuperscript{0}a\textsuperscript{0}), (D) tetragonal phase at 600 K (a\textsuperscript{0}a\textsuperscript{0}c\textsuperscript{+}). }
    \label{fig:cspbi3}
\end{figure}

\begin{figure}[H]
    \centering
    \includegraphics[scale=0.65]{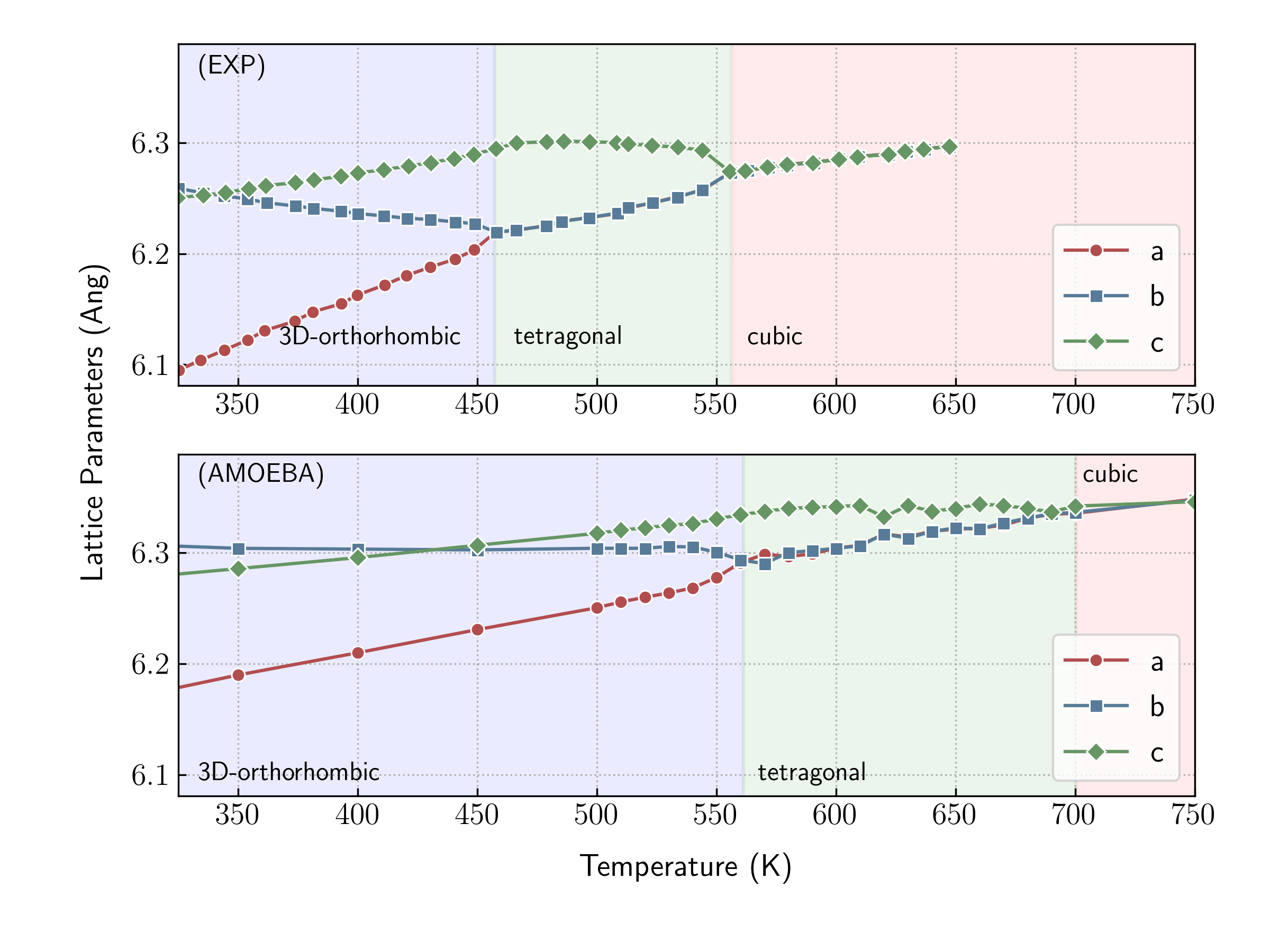}
    \caption{Pseudocubic lattice parameters for CsPbI\textsubscript{3} as a function of temperature, with phase boundaries indicated. Top: Experimental data\cite{marronnier2018anharmonicity}. Bottom: Simulation results with the AMOEBA force field.}
    \label{fig:cspbi3_phase_change}
\end{figure}

\begin{table}[H]
  \centering
  \begin{tabular}{c c c c}
    \hline
              & Experimental\cite{trots2008high, marronnier2018anharmonicity} & Ab-initio\cite{castelli2014bandgap} & AMOEBA \\
    \hline
    \multicolumn{4}{l}{Non Perovskite 1D-Orthorhombic}\\
    a          & 4.802 (298 K)\textsuperscript{a}       & 4.773      &  5.000  (100K) \\
    b          & 10.458       & 10.400     &  10.324 \\
    c          & 17.776       & 17.651     &  17.998 \\
    $\Delta E$ & -            & -0.385    &  -0.528 \\
    \hline
    \multicolumn{4}{l}{3D-Orthorhombic} \\
    a          & 8.620 (324.6 K)\textsuperscript{b}  & 8.372  & 8.812  (450 K)\\
    b          & 8.852          & 8.974  & 8.912         \\ 
    c          & 12.500         & 12.403 & 12.615        \\
    $\Delta E$ & -              & 0.00   & 0.00          \\
    \hline
    \multicolumn{4}{l}{Tetragonal} \\
    a          & 8.826 (511.7 K)\textsuperscript{b}  & 8.625    & 8.921 (600 K)  \\ 
    c          & 12.598  & 12.713   & 12.676 \\  
    $\Delta E$ & -  & 0.203   & 0.558  \\ 
    \hline
    \multicolumn{4}{l}{Cubic} \\
    a          & 6.297 (646.5 K)\textsuperscript{b}   & 6.275    & 6.336 (700 K) \\
    $\Delta E$ & -       & 0.864   & 1.000         \\
    \hline
  \end{tabular}
  \caption{Experimental, ab-initio and calculated AMOEBA lattice parameters (a, b, c, in \AA) and relative energies ($\Delta E$ in kcal/mol) of CsPbI\textsubscript{3} crystal polymorphs.}
  \label{tab:cspbi3_lattice}
\end{table}

We evaluated the ability of the AMOEBA force field to reproduce this phase behavior, and found that all four polymorphs were able to be stabilized, as shown in figure \ref{fig:cspbi3}. In addition, the order of stability of the polymorphs agrees with previous ab initio calculations \cite{castelli2014bandgap}, and the lattice parameters agree well with experimental values (see table \ref{tab:cspbi3_lattice}). Upon heating, the 1D-orthorhombic structure remained stable up to 900 K. While we did not observe a transition to the cubic structure, this is not surprising given the limited timescale accessible to direct simulation, as the transition involves large-scale rearrangement of atomic positions and is therefore likely to be a rare and slow event, as confirmed by recent work\cite{bischak2019liquid}. Heating of the 3D-orthorhombic structure did result in phase transitions to first the tetragonal structure (at 550 K) and then the cubic structure (at 690 K), as shown in figure \ref{fig:cspbi3_phase_change}. These transition temperatures are elevated relative to experiment, but the overall phase behavior agrees rather well, including the crossover in the 3D-orthorhombic lattice parameters.

\subsubsection{CH\textsubscript{3}NH\textsubscript{3}PbI\textsubscript{3}}
CH\textsubscript{3}NH\textsubscript{3}PbI\textsubscript{3} has been studied extensively. Experiments show a distinct 3D-orthorhombic phase below 160 K, (a\textsuperscript{-}a\textsuperscript{-}c\textsuperscript{+}) tilting pattern in Glazer notation\cite{glazer1972classification}, a cubic phase (a\textsuperscript{0}a\textsuperscript{0}a\textsuperscript{0}) above 330 K and a tetragonal phase (a\textsuperscript{0}a\textsuperscript{0}c\textsuperscript{-}) in between\cite{whitfield2016structures}. 

\begin{figure} [H]
    \centering
    \includegraphics[scale=0.6]{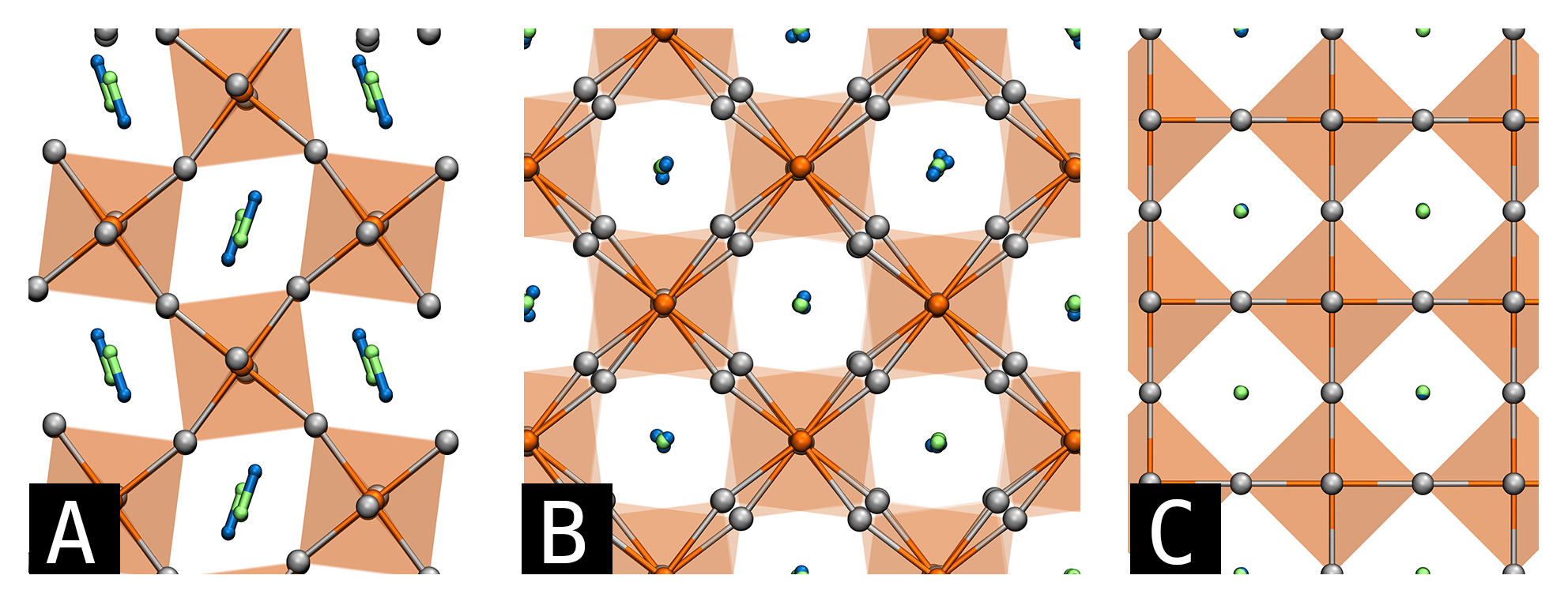}
    \caption{Averaged atomic positions of CH\textsubscript{3}NH\textsubscript{3}PbI\textsubscript{3} perovskite phases from AMOEBA simulations. Pb, I, C and N atoms are colored in orange, grey, green and blue, respectively. Hydrogens were hidden for clarity. (A) 3D-orthorhombic phase at 150 K (a\textsuperscript{-}a\textsuperscript{-}c\textsuperscript{+}), (B) tetragonal phase at 440 K (a\textsuperscript{0}a\textsuperscript{0}c\textsuperscript{-}), (C) cubic phase at 500 K (a\textsuperscript{0}a\textsuperscript{0}a\textsuperscript{0}). C and N atoms appear superimposed due to rotational averaging in both tetragonal (B) and cubic (C) phases.}
    \label{fig:mapbi3}
\end{figure}

\begin{figure} [H]
    \centering
    \includegraphics[scale=0.65]{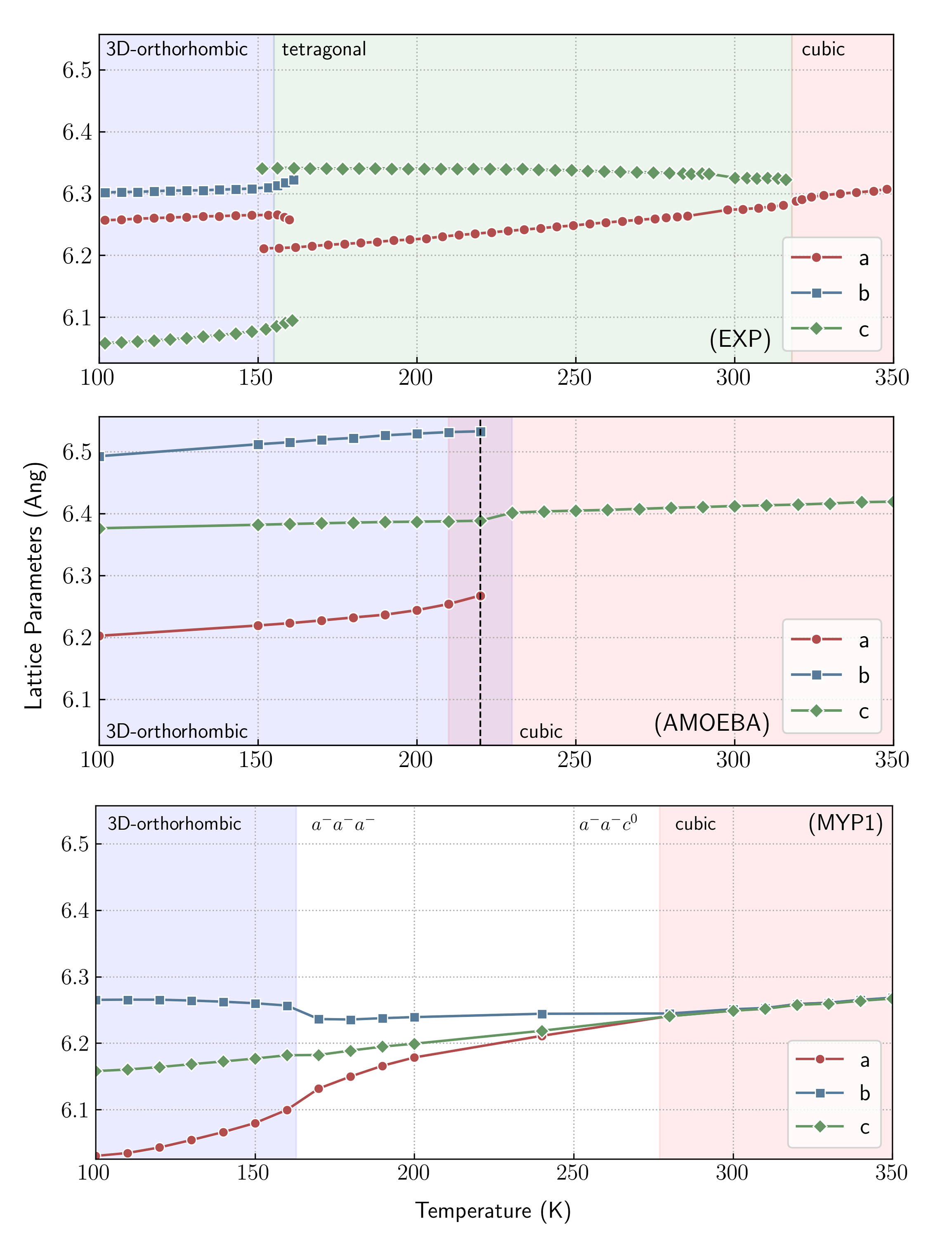}
    \caption{Pseudocubic lattice parameters for CH\textsubscript{3}NH\textsubscript{3}PbI\textsubscript{3} as a function of temperature. Top: Experimental data. Middle: Simulation results with the AMOEBA force field. The shaded area centered at the transition temperature of 220 K (dotted line) indicates the region of hysteresis from separate heating and cooling cycles. Bottom: Simulation results with the MYP1 point charge model.}
    \label{fig:mapbi3_phase_change}
\end{figure}

We evaluated the ability of the AMOEBA force field to reproduce this phase behaviour, and found that all three polymorphs were able to be stabilized, as shown in figure \ref{fig:mapbi3}. Stabilization of the tetragonal phase, however, required the introduction of hetero-atomic vdW parameters, which is discussed in more detail below. Using the unmodified AMOEBA potential, the phase transition from the 3D-orthorhombic to the cubic structure occurred at roughly 220 K, as shown in figure \ref{fig:mapbi3_phase_change}.
Heating and cooling cycles reveal a small region of hysteresis, with a transition at 230 K upon heating and at 210 K upon cooling. The discontinuity of the lattice parameters indicates a first-order phase transition. Experiments also show a first-order phase transition, though from the orthorhombic to the tetragonal phase\cite{sharada2016ch3nh3pbi3,jin2019characterizationof}, followed by a second-order phase transition (although debate is still ongoing) from the tetragonal to the cubic phase \cite{whitfield2016structures, weller2015complete, sharada2016ch3nh3pbi3,jin2019characterizationof}. 
With MYP1, on the other hand, we observed a continuous transition from the orthorhombic to the cubic phase, passing through tilting patterns (a\textsuperscript{-}a\textsuperscript{-}a\textsuperscript{-} and a\textsuperscript{-}a\textsuperscript{-}c\textsuperscript{0}) from 170-240 K that differ from the experimental tetragonal phase.

In addition to not stabilizing the tetragonal phase, our unmodified AMOEBA potential yields lattice parameters that are slightly larger than experimental values. Comparison with MYP1\cite{mattoni2015methylammonium} indicates that this is at least partly due to weaker tilting of the octahedra in the 3D-orthorhombic structure. We were able to improve the lattice parameters and stabilize the intermediate tetragonal phase by introducing hetero-atomic vdW parameters, as shown in table \ref{tab:mapbi3_lattice_params}. In particular, we found that reducing $R_{min}$ by 7 percent between Pb-CH\textsubscript{3}NH\textsubscript{3} and I-CH\textsubscript{3}NH\textsubscript{3} increased the octahedral tilting. This also gave the correct ordering of energies for the different polymorphs, although with a 3D-orthorhombic phase that was too stable relative to the others. As hetero-atomic vdW pairs are not supported in OpenMM, canonical Tinker\cite{rackers2018tinker} was used for these simulations. Modification of hetero-atomic vdW pairs compensate for charge transfer and penetration effects\cite{semrouni2013ab}, which AMOEBA does not include in the potential function. The introduction of hetero-atomic parameters could potentially lower transferability, but there are plans to include charge transfer and charge penetration explicitly in a future version of AMOEBA\cite{liu2019amoeba+}. 

\begin{table} [H]
  \centering
  \begin{tabular}{c c c c c}
    \hline
              & Experimental\cite{whitfield2016structures} & Ab-initio\cite{ong2015structural, yin2014anomalous} & AMOEBA\textsuperscript{a}  & AMOEBA\textsuperscript{b}\\
    \hline
    \multicolumn{5}{l}{3D-Orthorhombic phase} \\
    a          & 8.560 (10 K)   & 8.840    & 9.210 (150 K) & 8.693 (160 K)\\
    b          & 8.812          & 8.770    & 8.790         & 8.811      \\
    c          & 12.587         & 12.970   & 12.760        & 12.714      \\
    $\Delta E$ & -              & 0.00     & 0.0           & 0.0       \\
    \hline
   \multicolumn{5}{l}{Tetragonal phase} \\
    a          & 8.799 (190 K)   & 8.80  & -  & 8.952 (440 K)  \\
    c          & 12.688          & 12.99 & -  & 12.734         \\
    $\Delta E$ & -               & 0.55 & -  & 1.286             \\
    \hline
    \multicolumn{5}{l}{Cubic phase} \\
    a          & 6.306 (350 K) & 6.39   & 6.433 (450 K) & 6.351 (500 K) \\
    $\Delta E$ & -             & 0.416   & 0.993         & 1.480 \\
    \hline
  \end{tabular}
  \caption{Experimental, ab-initio and AMOEBA lattice parameters (a,b,c in \AA) and relative energies ($\Delta E$ in kcal/mol) of CH\textsubscript{3}NH\textsubscript{3}PbI\textsubscript{3} crystal polymorphs. (\textsuperscript{a} Without hetero-atomic vdW pairs. \textsuperscript{b} With Pb-CH\textsubscript{3}NH\textsubscript{3} and I-CH\textsubscript{3}NH\textsubscript{3} hetero-atomic $R_{min}$ reduced by 7\%.) }
  \label{tab:mapbi3_lattice_params}
\end{table}

\subsection{Simulation Code Availability and Benchmarking}
Unlike point charge models, AMOEBA is only available in two simulation packages, OpenMM and Tinker. This is likely due to its complexity which requires extensive modification of typical core routines. The Tinker software family includes canonical Tinker with OpenMP support, Tinker-OpenMM with GPU support and Tinker-HP with MPI support. As these codes were mainly developed with biological systems in mind, they do not include built-in functions to extract or calculate properties such as the pressure tensor and Young's modulus. However the python interface and a wide variety of external packages add customization and rapid prototyping capabilities to OpenMM. 

We compared the efficiency of point charge models and the AMOEBA force field using several different  codes to simulate CH\textsubscript{3}NH\textsubscript{3}PbI\textsubscript{3}. MPI and GPU versions of LAMMPS were used to benchmark point charge models. OpenMM and Tinker HP were used to benchmark the AMOEBA force field utilizing GPU and MPI capabilities, respectively. In addition, a simple GAFF point charge model was tested in OpenMM with the GPU capability for a fair comparison. Our results are summarized in table \ref{tab:benchmark}.

\begin{table}
    \centering
    \begin{tabular}{c c c c }
    \hline
    \multicolumn{3}{c}{Benchmark (ns/day)} \\
    \hline
    \multicolumn{3}{c}{MPI} \\
    \hline
     Number of cores  & LAMMPS  & TinkerHP  \\
     \hline
     1               &  1.000   & 0.041     \\
     4               &  4.010   & 0.144     \\
     8               &  7.541   & 0.244     \\
     16              & 12.637   & 0.419     \\
     32              & 25.994   & -         \\
    \hline
    \multicolumn{3}{c}{GPU} \\
    \hline
    OpenMM  & GAFF & AMOEBA  \\
    \hline
    NVIDIA V100 SXM2 & 174.351 & 12.273 \\
    \hline
    \end{tabular}
    \caption{MPI benchmark (ns/day) of LAMMPS (MYP1) and TinkerHP v1.0 (AMOEBA) of an orthorhombic  CH\textsubscript{3}NH\textsubscript{3}PbI\textsubscript{3} perovskite with 5184 atoms with a timestep of 0.5fs. The same system was used in GPU comparison in OpenMM with GAFF and AMOEBA.}
    \label{tab:benchmark}
\end{table}

Overall, we find that the simple point charge models run 14-30 times faster than AMOEBA on GPUs and CPUs respectively, mainly due to AMOEBA requiring the iterative calculation of induced dipoles for each site until convergence. This adds a considerable computational cost to the simulation, which can be especially severe for high-valent ions\cite{shi2013polarizable}. We note that the use of the iAMOEBA water model (inexpensive AMOEBA) would improve these numbers by calculating the induced dipoles directly from the permanent multipoles, removing the need for the iterative process described above\cite{wang2013systematic}. Nonetheless, using a single V100 GPU, we could simulate 12 ns/day, which indicates that studies of crystal nucleation and growth should be feasible up to the microsecond timescale by direct simulation and beyond using advanced sampling methods.

\section{Conclusion}
In this paper, we tested the ability of the AMOEBA polarizable force field to describe the solid-state and solution properties of metal halide perovskites and their precursors and compared it with several non-polarizable force fields from the literature. AMOEBA simulations show good agreement with experiments for both solid and solution state properties for both CsPbI\textsubscript{3} and CH\textsubscript{3}NH\textsubscript{3}PbI\textsubscript{3} perovskites in spite of minimal fitting effort (only lead and methylammonium ions were parameterised by us). Our results indicate that the AMOEBA force field is promising for studying a wide range of metal halide perovskites and their precursors in complex environments, including solid-liquid interfaces.

\section{Supplementary Material}
The new AMOEBA parameters (Pb\textsuperscript{2+} and CH\textsubscript{3}NH\textsubscript{3}\textsuperscript{+}) and detailed visualization of orthorhombic and tetragonal phases of CH\textsubscript{3}NH\textsubscript{3}PbI\textsubscript{3} can be found in the supplementary material.

\section{Acknowledgement}
The authors thank the Australian Research Council (ARC) and the Centre of Excellence in Exciton Science for financial assistance (Grant no. CE170100026), and the Sydney Informatics Hub (SIH), the University of Sydney's high performance computing cluster "Artemis", NCI's "Raijin" high performance computing cluster and Pawsey's "Magnus" high performance computing cluster for providing the computing resources that have contributed to the results reported herein.


%
%

%


\newpage
\newpage
\bibliography{references.bib}

\end{document}


\maketitle
\thispagestyle{empty}

\lstset{ basicstyle=\ttfamily\scriptsize,breaklines=true}

\newpage
\section{AMOEBA parameters}
\subsection{Pb2+}
\begin{lstlisting}
atom        888  888    Pb+   "Lead Ion Pb++"               82   207.200    0
vdw         888               3.600     0.4000
multipole   888    0    0               2.00000
                                        0.00000    0.00000    0.00000
                                        0.00000
                                        0.00000    0.00000
                                        0.00000    0.00000    0.00000
polarize    888          2.0424     0.1500
\end{lstlisting}

\subsection{CH3HN3+}
\begin{lstlisting}
atom          801    801    C     "MA+ C               "         6    12.011    4
atom          802    802    H     "MA+ CH              "         1     1.008    1
atom          803    803    N     "MA+ N               "         7    14.007    4
atom          804    804    H     "MA+ NH              "         1     1.008    1

# Estimated van der Waals Parameters :
 vdw         801               3.820     0.1010
 vdw         802               2.780     0.0260     0.91
 vdw         803               3.710     0.1050
 vdw         804               2.700     0.0200     0.91

# Estimated Bond Stretching Parameters :
 bond        801  802          400.0     1.0880
 bond        801  803          400.0     1.5077
 bond        803  804          520.0     1.0243

# Estimated Angle Bending Parameters :
 angle       802  801  802     34.50     110.83
 angle       802  801  803     50.60     108.08
 angle       801  803  804     35.00     111.48
 angle       804  803  804     35.00     107.39

# Estimated Stretch-Bend Parameters :
 strbnd      802  801  803     11.50      18.70
 strbnd      801  803  804      7.20       4.30
 ureybrad      0    0    0       0.0     0.0000
 ureybrad      0    0    0       0.0     0.0000

# Estimated Torsional Parameters :
torsion     802  801  803  804      0.000 0.0 1   1.000 180.0 2   0.500 0.0 3

# Multipole Parameters
multipole   801  803  802              -0.02971
                                        0.00000    0.00000    0.29620
                                       -0.51434
                                        0.00000   -0.51434
                                        0.00000    0.00000    1.02868
multipole   802  801  803               0.10641
                                       -0.02935    0.00000   -0.08013
                                       -0.06588
                                        0.00000   -0.00219
                                       -0.00460    0.00000    0.06807
multipole   803  801  804               0.13244
                                        0.00000    0.00000    0.07867
                                       -0.18214
                                        0.00000   -0.18214
                                        0.00000    0.00000    0.36428
multipole   804  803  801               0.19268
                                        0.03265    0.00000   -0.15362
                                       -0.00279
                                        0.00000   -0.00443
                                        0.03642    0.00000    0.00722

polarize     801         1.3340     0.3900     802  803 
polarize     802         0.4960     0.3900     801 
polarize     803         1.0730     0.3900     801  804
polarize     804         0.4960     0.3900     803 
\end{lstlisting}

\newpage

\subsection{Orthorhombic and tetragonal phases}
\subsubsection{CsPbI\textsubscript{3}}

\begin{figure}[h]
    \centering
    \includegraphics[scale=0.85]{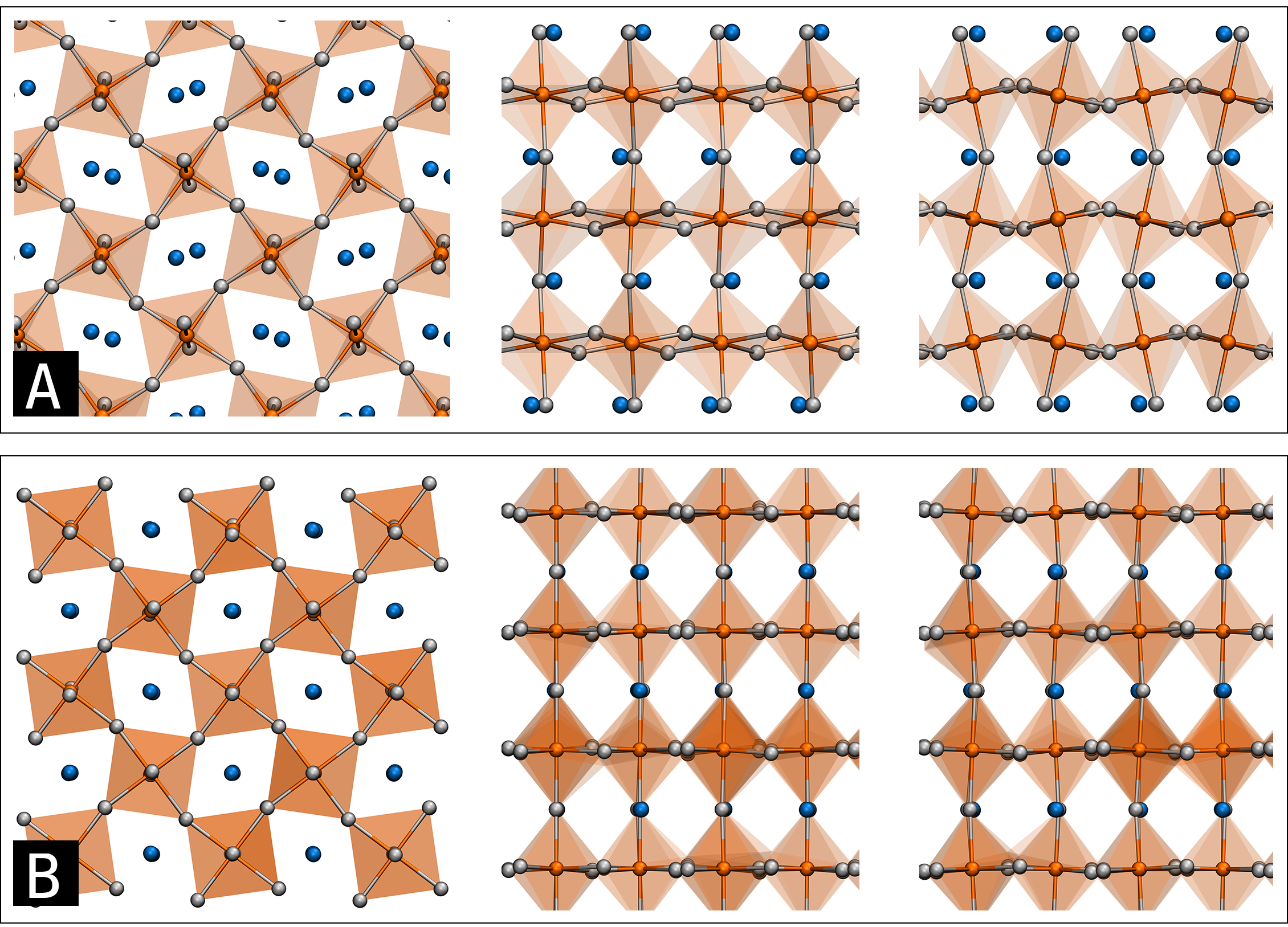}
    \caption{Averaged structures from AMOEBA simulations.  A) 3D-orthorhombic phase (450 K). From left, 001, 100 and 010 planes. B) Tertagonal phase (600 K). From left, 001, 100, and 101 planes}
    \label{fig:cspbi3}
\end{figure}

\newpage
\subsubsection{CH\textsubscript{3}NH\textsubscript{3}PbI\textsubscript{3}}

\begin{figure}[h]
    \centering
    \includegraphics{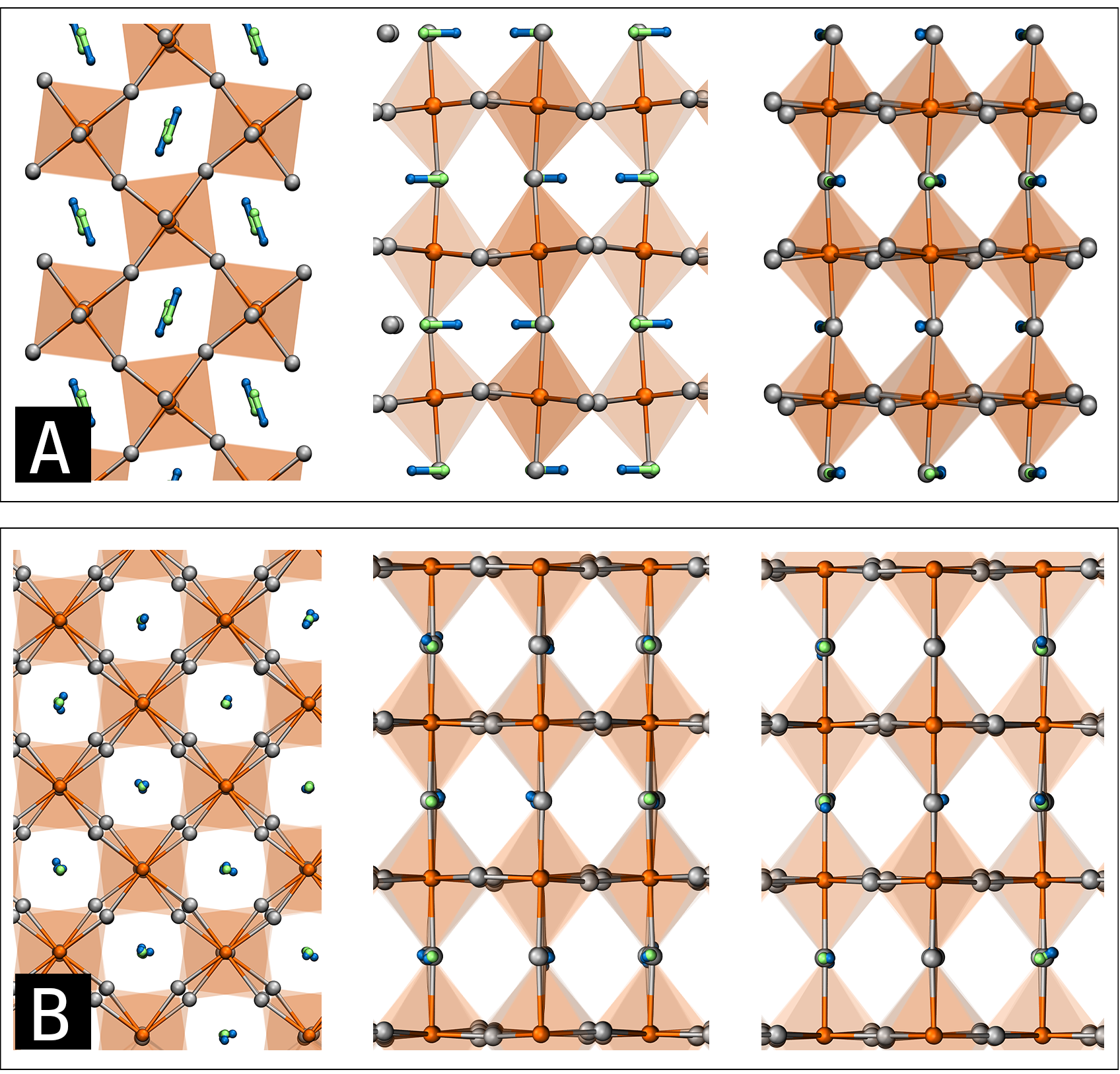}
    \caption{Averaged structures from AMOEBA simulations.  A) 3D-orthorhombic phase (150 K). From left, 001, 100 and 010 planes. B) Tertagonal phase (440 K) with reduced hetero-atomic $R_{min}$ between Pb-CH\textsubscript{3}NH\textsubscript{3}PbI\textsubscript{3} and I-CH\textsubscript{3}NH\textsubscript{3}PbI\textsubscript{3} by 7 percent. From left, 001, 100, and 101 planes. C and N atoms look superimposed due to rotational averaging in tetragonal (B) phase. H atoms were hidden for clarity.}
    \label{fig:mapbi3}
\end{figure}
